\documentclass{PoS}

\usepackage[T1]{fontenc}
\usepackage[utf8]{inputenc}
\usepackage{amsmath}
\usepackage{amssymb}
\usepackage{graphicx}  
\usepackage{epsfig}

\title{Polyakov loop correlators and cyclic Wilson loop from lattice QCD}

\ShortTitle{Polyakov loop correlators and cyclic Wilson loop from lattice QCD}

\author{Alexei Bazavov\\
       University of California, Riverside/University of Iowa
       Department of Physics and Astronomy, Iowa City, Iowa 52242-1479, USA\\
       E-mail: \email{obazavov@quark.phy.bnl.gov}}
\author{Peter Petreczky\\
       Brookhaven National Laboratory, Physics Department, Upton NY 11973-5000, USA\\
       E-mail: \email{petreczk@quark.phy.bnl.gov}}
\author{\speaker{Johannes Heinrich Weber}
\\
        Technische Universit\"{a}t M\"{u}nchen, Physik Department T30f, James-Franck-Str. 1, 85748 Garching, Germany\\
        E-mail: \email{johannes.weber@tum.de}}

\author{(TUMQCD Collaboration)}

\abstract{We discuss 
color screening in 2+1 flavor QCD in terms of free energies of a static quark-antiquark pair. 
Thermal modifications of long distance correlations in quark-antiquark systems are studied in terms of static meson correlators. 
We calculate the Polyakov loop correlator, the color-singlet Wilson line correlator in Coulomb gauge and the cyclic Wilson loop on lattices using the HISQ/Tree action and almost physical quark mases with $ N_\tau=4,6,8,10,12 $. We present results in the continuum limit for temperatures up to $ T \lesssim 650\,\mathrm{MeV} $ and discuss the linear divergence of the cyclic Wilson loop.}

\FullConference{The 33rd International Symposium on Lattice Field Theory\\
		14 -18 July 2015\\
		Kobe International Conference Center, Kobe, Japan*}

\begin{document}

\section{Introduction}\label{intro}
\vskip-1ex

QCD is known to exhibit a confined phase at low temperatures and a deconfined phase at high temperatures. 
In the confined phase, quarks and gluons are bound inside of hadrons and the iso-vector chiral symmetry in the light quark sector is broken spontaneously. 
This spontaneously broken chiral symmetry is restored in the deconfined phase. 

In $ SU(N_c) $ pure gauge theory without dynamical quarks, the deconfinement transition is a phase transition, with the Polyakov loop being the order parameter. 
This phase transition is associated with the $ Z(N_c) $ center symmetry. The logarithm of the Polyakov loop corresponds to the negative free energy of a static quark in temperature units~\cite{McLerran:1981pb}. 
The Polyakov loop is not invariant under $ Z(N_c) $ transformation and therefore strictly vanishes for temperatures below the phase transition temperature, \mbox{i.e.} the free energy of a static quark is infinite. 
On the other hand, it assumes a finite expectation value above the transition temperature due to color screening. 
Thus, the deconfinement transition is closely related to the onset of color screening~\cite{McLerran:1981pb}.

The center symmetry is broken explicitly by the presence of sea quarks in full QCD.  
As two color charges are separated beyond the string breaking distance, the flux tube breaks apart and a light quark-antiquark pair is created from the sea. 
Hence, the Polyakov loop takes a small, but positive expectation value in the confined phase of QCD, which is related to the binding energy of static-light mesons. 
Therefore, it is not clear to what extent the Polyakov loop is suitable for the discussion of the deconfinement transition in QCD, and therefore, deconfinement in QCD is often discussed in terms of fluctuations of conserved charges (see~\cite{Mukherjee:2015mxc} and references therein).

In this contribution, we discuss color screening in QCD with $ N_c=3 $ in terms of the Polyakov loop correlator or equivalently the free energy $ F_{Q\bar Q}(r,T) = F_{\mathrm{avg}}(r,T) $ of a static $ Q \bar Q $ pair. We also consider a color-singlet Wilson line correlator and the cyclic Wilson loop, which are both related to the color-singlet and color-octet free energies $ F_S(r,T) $ and $ F_O(r,T) $ for a $ Q \bar Q $ pair in full QCD. 
These quantities have already been studied in \mbox{Ref.}~\cite{Bazavov:2013zha}, though the discussion was restricted to $ N_\tau=6 $ lattices. 
Here, we include finer lattices and extrapolate to the continuum limit. 

\section{Lattice setup}
\vskip-1ex

We study the Polyakov loop and the related correlators in full QCD for a large range of temperatures and five different temporal extents $ N_\tau=\{12,10,8,6,4\} $ of the lattice. The temperature $ T(\beta,N_\tau) =1/(a(\beta)N_\tau) $ is related to the gauge coupling $ \beta=10/g^2 $ through the lattice spacing $ a(\beta) $. Taking the continuum limit at fixed temperature requires simultaneous variation of both $ \beta $ and $ N_\tau $, which is why numerical studies of QCD at finite temperature are quite demanding. In order to avoid finite volume effects, we use lattices with aspect ratio $ N_\sigma/N_\tau=4 $, where $ N_\sigma $ is the extent of each spatial direction. 
We have a physical strange quark mass and nearly physical light quark masses, $ m_l=m_s/20 $, corresponding to a pion mass of $ m_\pi=161\,\mathrm{MeV} $ in the continuum limit. The gauge configurations were generated with tree-level improved Symanzik gauge action and highly improved staggered quarks (HISQ) by the HotQCD collaboration~\cite{Bazavov:2011nk,Bazavov:2014pvz}. 
We use the publicly available MILC code~\cite{code} in combination with libraries provided by the USCQD consortium~\cite{usqcd}.

\enlargethispage{1\baselineskip}

\subsection{Observables}

The Polyakov loop $ L(\beta,N_\tau,\mathbf{x}) $ is defined in lattice QCD as a normalized trace of a temporal Wilson line $ W(\beta,N_\tau,\mathbf{x}) $ wrapping around the time direction once,
\begin{equation}
  L(\beta,N_\tau,\mathbf{x})   = \frac{1}{3}\, \mathrm{Tr}\, W(\beta,N_\tau,\mathbf{x}), \quad  W(\beta,N_\tau,\mathbf{x}) = \prod_{x_0=1}^{N_\tau} U_0(x_0,\mathbf{x}).
\end{equation}\vskip-1ex
We obtain the expectation values of Polyakov loop, Polyakov loop correlator, Wilson line correlator and cyclic Wilson loop by taking ensemble averages (average over $ \mathbf{x} $ included) as 
\begin{align}
  L^{\mathrm{bare}}(\beta,N_\tau) =&\  \langle L(\beta,N_\tau,\mathbf{x}) \rangle, \\
  C_P^{\mathrm{bare}}(\beta,N_\tau,r) =&\ \langle L(\beta,N_\tau,\mathbf{x}) L^\dagger(\beta,N_\tau,\mathbf{x}+\mathbf{r}) \rangle, \\
  C_S^{\mathrm{bare}}(\beta,N_\tau,r) =&\ {1}/{3}\, \langle \mathrm{Tr}\, \left[ W(\beta,N_\tau,\mathbf{x}) W^\dagger(\beta,N_\tau,\mathbf{x}+\mathbf{r}) \right] \rangle, \label{eq: CSbare}\\
  W_S^{\mathrm{bare}}(\beta,N_\tau,r) =&\ {1}/{3}\, \langle \mathrm{Tr}\, \left[ W(\beta,N_\tau,\mathbf{x}) S(\beta,N_\tau,\mathbf{x};\mathbf{r}) W^\dagger(\beta,N_\tau,\mathbf{x}+\mathbf{r})  S^\dagger(\beta,0,\mathbf{x};\mathbf{r}) \right] \rangle  \label{eq: WSbare}.
\end{align}\vskip-1.0ex
Since the Wilson line $ W(\beta,N_\tau,\mathbf{x}) $ itself is not gauge invariant, $ C_S^{\mathrm{bare}}(\beta,N_\tau,r) $ vanishes unless a gauge-fixing procedure is used.
We fix the gauge fields to Coulomb gauge~\cite{Kaczmarek:2002mc,Digal:2003jc}. 
In the definition of $ W_S^{\mathrm{bare}}(\beta,N_\tau,r) $, we use spatial Wilson lines $ S(\beta,x_0,\mathbf{x};\mathbf{r}) $ that are path ordered products of link matrices along the shortest paths from point $ \mathbf{x} $ to point $ \mathbf{x}+\mathbf{r} $ within time slice $ x_0 $. 
We only consider paths aligned with the lattice axes, since paths of different shape would introduce additional path-dependent cusp and intersection divergences. 
The spatial Wilson lines ensure gauge invariance such that gauge fixing is not required for cyclic Wilson loops.
 
All four bare observables diverge in the continuum limit and require multiplicative renormalization with a factor $ \exp{[-N_\tau c_Q(\beta)]} $ for each Wilson line $ W(x) $. 
Moreover, spatial Wilson lines in the cyclic Wilson loop introduce a linear divergence. We regulate the Wilson loop by applying link smearing procedures to the spatial links,  
using $ n_{\mathrm{hyp}}=\{0,1,2,5\} $ iterations of HYP smearing restricted to spatial links only. 
After taking care of the divergence of the Wilson line $ W(x) $, the renormalized observables are related to free energies for finite $ N_\tau $ and in the continuum limit as
\begin{equation}
  L^{\mathrm{ren}}(T) = e^{-F_Q(T)/T}\quad , \quad
  C_P^{\mathrm{ren}}(T,r) = e^{-F_{\mathrm{avg}}(r,T)/T}\quad \text{and} \quad 
  C_S^{\mathrm{ren}}(T,r) = e^{-F_S(r,T)/T}.
\end{equation}\vskip-1ex
The cyclic Wilson loop and the Wilson line correlator both mix singlet and octet free energies under renormalization. 
Nevertheless, the differences with the Polyakov loop correlator,
\begin{equation}
  [C_S-C_P](T,r) = 8/9\, [C_S-C_O](T,r), \qquad [W_S-C_P](T,r) = 8/9\, [W_S-W_O](T,r),
  \label{eq: multiplicative renormalization}
\end{equation}
are multiplicatively renormalizable~\cite{Berwein:2013xza}. 
The octet Wilson line correlators $ C_O $ and octet Wilson loops $ W_O $ are defined analogously to $ C_S $ and $ W_S $ in \mbox{eqs.}~(\ref{eq: CSbare}) and~(\ref{eq: WSbare}) with a pair of $ SU(3) $ generators $ T^a $ left- and right-multiplied to one of the two temporal Wilson lines $ W(\beta,N_\tau,\mathbf{x}) $ and summation over the group index $ a $. 
In the following discussion, we make use of free energies in units of the temperature, \mbox{e.g.} $ f_Q(T,N_\tau) =F_Q(T,N_\tau)/T $, which vary more slowly with the temperature than the primary observables and are renormalized by adding $ N_\tau c_Q(\beta) $ (or the respective combinations for the other free energies). We also usually omit the indication $ ^{\mathrm{ren}} $ for renormalized quantities.

\subsection{Renormalization and scale setting with $ Q\bar Q $ procedure}

We obtain the renormalization constant $ c_Q(\beta) $ by normalizing the $ T=0 $ static $ Q\bar Q $ potential $ V(r) $ to a prescribed value. 
Namely, $ V(r) $ for different lattice spacings $ a(\beta) $ is fixed to be $ 0.954/r_0 $ or $ 0.2065/r_1 $ at distances $ r=r_0 $~\cite{Bazavov:2011nk} or $ r=r_1 $~\cite{Bazavov:2014pvz}, respectively.
The scales $ r_0 $ and $ r_1 $ are defined as
\begin{equation}
  r^2 \left.\frac{d V(r)}{d r}\right|_{r=r_i}= C_i, \quad i=0,1,
  \label{eq: QQbar procedure}
\end{equation}\vskip-1ex
where $ C_0=1.65 $ and $ C_1=1.0 $. 
In physical units we have $ r_0=0.4688(41)\,\mathrm{fm} $ and $ r_1=0.3106\,\mathrm{fm} $~\cite{Bazavov:2010hj}. 
Either distance $ r_i $ defines the lattice spacing $ a(\beta) $ in physical units as a function of the gauge coupling $ \beta $. 
The renormalization scheme ($ Q\bar Q $ procedure) is limited by availability of $ T=0 $  lattice data ($ \beta \leq 7.825 $, lattice spacing $ a\gtrsim 0.04\,\mathrm{fm} $). 
Thus, continuum extrapolation using the finest lattices ($ N_\tau=12 $) is limited to $ T \lesssim 407\,\mathrm{MeV} $.
The renormalization constants $ c_Q(\beta) $ have been obtained using results on $ V(r) $ from~\cite{Bazavov:2011nk,Bazavov:2014pvz}. 
Because we need intermediate $ \beta $ values for a common set of temperatures, we interpolate $ c_Q(\beta) $ using R statistical package~\cite{Rscript} with error propagation via bootstrap method. 
Further details are deferred to a publication in preparation~\cite{Preprint}.

\section{Color screening observables from the static quark correlators}
\vskip-1ex

On the one hand, any static quark-antiquark free energy approaches the same constant at sufficiently large distances due to color screening. 
This constant equals twice the free energy of an isolated static quark because the two color charges are fully decorrelated. 
In other words the correlators defined above will approach $ L^2 $ at large distances. 
On the other hand, thermal effects are no more than a perturbation to correlators at sufficiently short distances. 
At distances, $ r \lesssim 1/T $ we expect to see the interplay between vacuum effects and medium effects, while at still larger distances we should see the onset of color screening. 
It is convenient to subtract the known asymptotic constant and obtain correlators that vanish for sufficiently large distances. 
By subtracting the bare Polyakov loop squared from the correlation functions the correlated noise at large distances is greatly reduced. 
The renormalized correlators are obtained by adding back the asymptotic constant in terms of the renormalized instead of the bare Polyakov loops.


\subsection{Polyakov loop and free energy $ F_Q(T) $}\label{sec: Polyakov loop and free energy F_Q(T)}
\vskip-1ex


We use bare Polyakov loop data from HotQCD given in~\cite{Bazavov:2011nk,Bazavov:2014pvz} 
and interpolate $ f_Q^{\mathrm{bare}}(\beta,N_\tau) $ for each $ N_\tau $ in $ \beta $ using R statistical package~\cite{Rscript} and error propagation via bootstrap method. 
Further details are deferred to a publication in preparation~\cite{Preprint}.
Interpolation is necessary to obtain the Polyakov loop for a common set of temperatures for all $ N_\tau $. 
We add the errors of $ N_\tau c_Q(\beta) $ in quadrature, since they are statistically independent from the errors of the bare Polyakov loop. 
%
We extrapolate $ F_Q(T,N_\tau) $ to the continuum limit with pointwise extrapolations for each temperature and with a global fit using a polynomial Ansatz,
\begin{equation}
  F_Q(T,N_\tau) = \sum_{i_0=0}^{N_0} a_{i_0} T^{i_0} + \frac{1}{N_\tau^2}\sum_{i_2=0}^{N_2} a_{i_2} T^{i_2}.
  \label{eq: global fit for F_Q}
\end{equation}
The orders $ N_0 $ and $ N_2 $ of two polynomials in $ T $ parametrize the temperature dependence of the continuum limit and of cutoff effects. 
Further details are deferred to a publication in preparation~\cite{Preprint}.
Our final continuum result, 
has been discussed already in~\cite{Petreczky:2015yta} with regard to comparison between $ Q \bar Q $ procedure and gradient flow for renormalizing the Polyakov loop, which yield a consistent continuum limit up to $ T \lesssim 400\,\mathrm{MeV} $.

\subsection{Color-singlet free energy $ F_S(r,T) $}\label{sec: Color-singlet free energy F_S(r,T) }

\begin{figure}
\centering
  \begin{minipage}{0.48\textwidth}
\centering
 \includegraphics[height=4cm,clip]{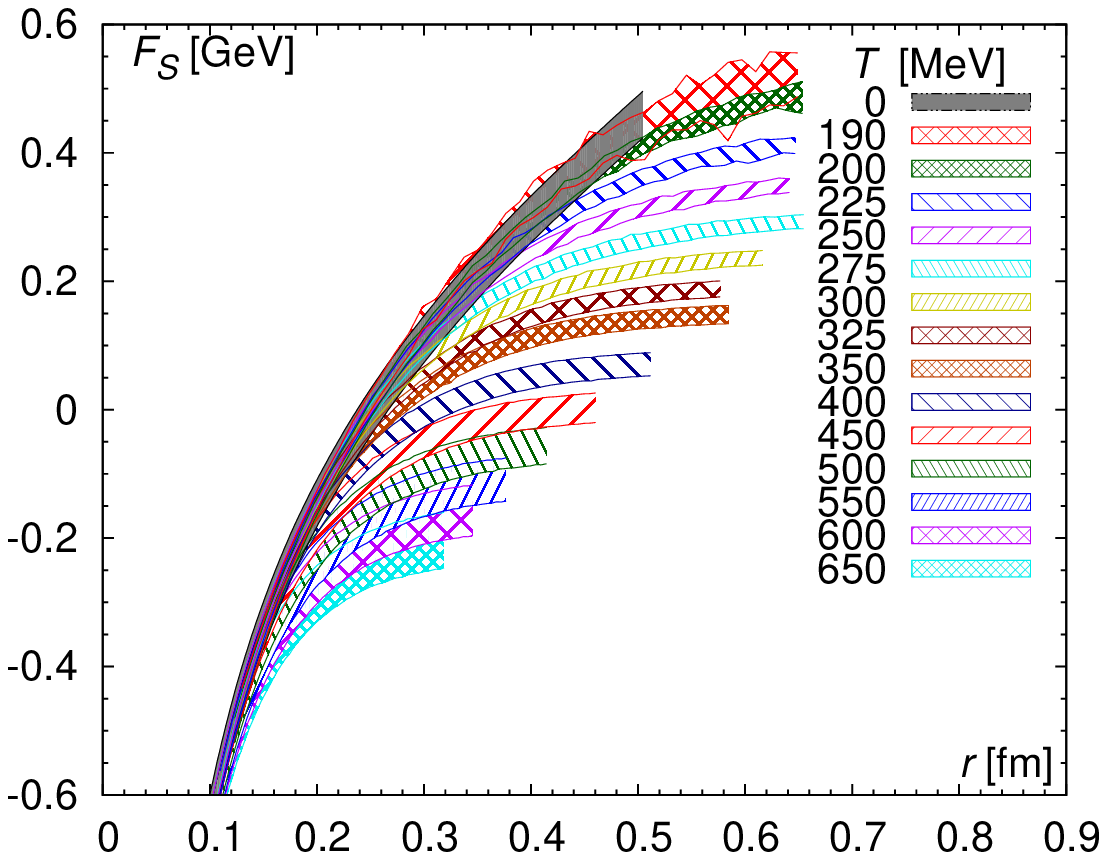}
  \vskip1ex
  \caption{\label{fig: Fscont}
  The singlet free energy $ F_S(r,T) $ is numerically similar to the zero temperature static energy $ V(r) $ up to $ rT \lesssim 0.45 $.
  }
\end{minipage}\hfill
\begin{minipage}{0.48\textwidth}
\centering
\includegraphics[height=4cm,clip]{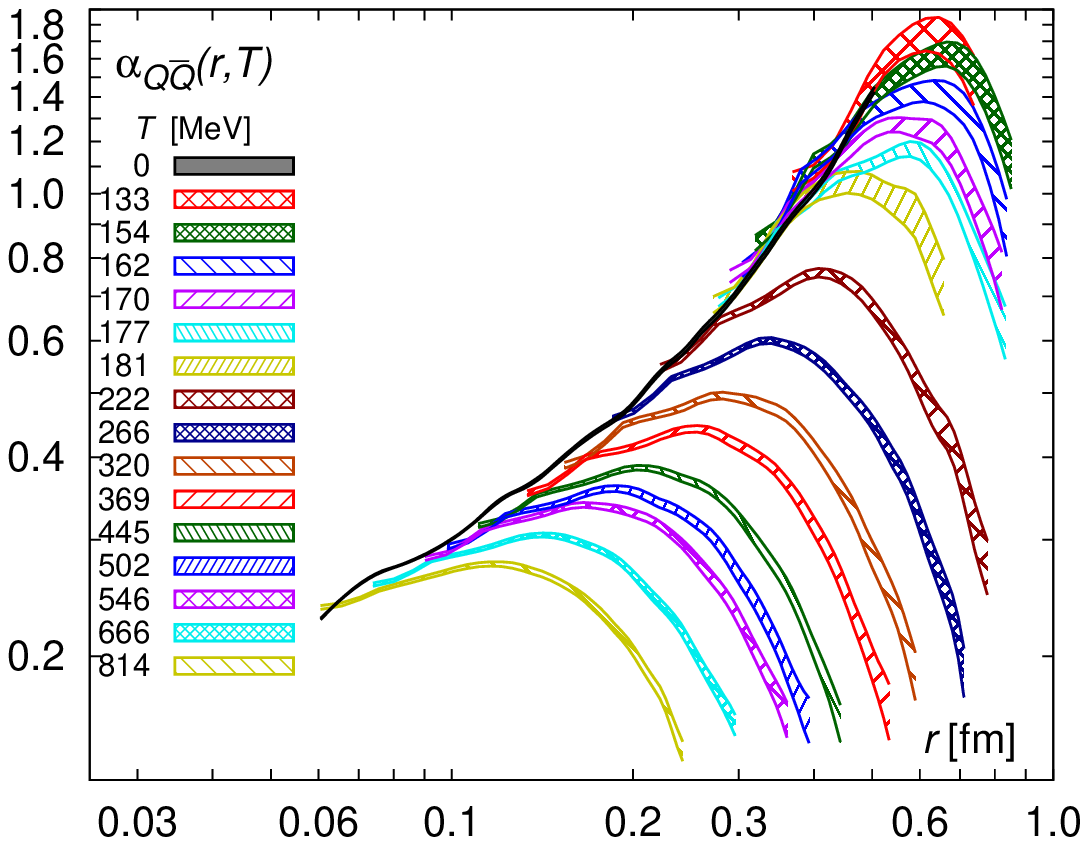}
  \caption{\label{fig: aeffnt6}
  The effective coupling $ \alpha_{Q \bar Q}(r,T,6) $ highlights the transition between confining and screening regimes. 
  The quadratic rise of the black band for $ T=0 $ indicates the string tension.
  }
\end{minipage}
\end{figure}

Applying the previously discussed steps to the color-singlet Wilson line correlator $ C_S(r,T,N_\tau) $, we compute $ f_S^{\mathrm{sub}}(r,T,N_\tau)=f_S(r,\beta,N_\tau)-2f_Q^{\mathrm{bare}}(\beta,N_\tau) $, the subtracted color singlet free energy.
As for the Poylakov loop, interpolation in the temperature is required for continuum extrapolation at fixed temperature.
Moreover, the available lattice distances $ r/a(\beta) $ correspond to different physical lengths for different $ \beta $. 
We correct for leading cutoff effects of the tree-level gauge action at short distances by using the improved distance $ r_i $ (see \mbox{e.g.}~\cite{Bazavov:2014soa}) instead of bare distance $ r_b $ and interpolate in the distance.
We perform both pointwise continuum extrapolation for each temperature and distance and with a global fit using a polynomial Ansatz similar to \mbox{eq.}~(\ref{eq: global fit for F_Q}) in both the temperature and the distance. 
Finally we add the continuum limit of the renormalized asymptotic value, $ 2f_Q(T) $, and obtain the renormalized singlet free energy. 
The continuum limit of this result is shown together with the static energy at zero temperature, $ V(r) $, in figure~\ref{fig: Fscont}. 
Both quantities are numerically similar up to $ rT \lesssim 0.45 $. This indicates a regime of almost vacuum-like physics with medium effects suppressed for small $ rT $. 
We define an effective coupling constant $ \alpha_{Q\bar Q}(r,T) $,
\begin{equation}
  \alpha_{Q\bar Q}(r,T,N_\tau) = -\frac{3}{4} r^2 \frac{\partial E(r,T,N_\tau)}{\partial r}, \quad E(r,T,N_\tau) = \{ F_S(r,T,N_\tau), V(r),\ldots \},
\end{equation}\vskip-1ex\noindent
to make these two regimes even more explict. 
$ \alpha_{Q\bar Q}(r,T,N_\tau) $ has only mild $ N_\tau $ dependence. We show show $ \alpha_{Q\bar Q}(r,T,6) $ together with $ \alpha_{Q\bar Q}(r,0) $ obtained from the $ T=0 $ static energy $ V(r) $ in figure~\ref{fig: aeffnt6}. 
The maximum defines a distance $ r_{\max}(T) $ where screening overcomes the string tension. 
Thus, we can identify vacuum-like behavior for $ r < r_{\max}(T) $ and the onset of color screening for $ r > r_{\max}(T) $.


\subsection{Color-averaged free energy $ F_{\mathrm{avg}}(r,T) $}\label{sec: Color-averaged free energy F_{avg}(r,T) }

\begin{figure}
\centering
\begin{minipage}{0.48\textwidth}
\centering
 \includegraphics[height=4cm,clip]{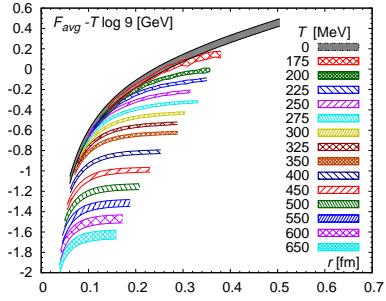}
\end{minipage} 
\begin{minipage}{0.48\textwidth}
\centering
  \caption{\label{fig: Facont}
  Once a trivial color factor $ T \log 9 $ is subtracted from the color-averaged free energy $ F_{\mathrm{avg}}(r,T) $ it is numerically similar to the zero temperature static energy $ V(r) $ up to $ rT \lesssim 0.15 $.
  }
\end{minipage}
\end{figure}

 
We compute $ f_{\mathrm{avg}}^{\mathrm{sub}}(r,T,N_\tau)=f_{\mathrm{avg}}(r,\beta,N_\tau)-2f_Q^{\mathrm{bare}}(\beta,N_\tau) $, the subtracted color-averaged free energy, 
by applying the same machinery to the Polyakov loop correlator $ C_P(r,T,N_\tau) $. 
It can be written in terms of the color-singlet free energy and the color-octet free energy\footnote{
The decomposition of the $ Q \bar Q $ free energy into singlet and octet contributions can be rigorously derived in the small distance limit using effective field theory approach~\cite{Brambilla:2010xn}.
},
\begin{equation}
  \exp{[-f_{\mathrm{avg}}(r,\beta,N_\tau)]} = {1}/{9}\ \exp{[-f_S(r,\beta,N_\tau)]} +{8}/{9}\ \exp{[-f_O(r,\beta,N_\tau)]}.
\end{equation}\vskip-1ex
Hadronic states with a $ Q \bar Q $ pair in color-octet configuration require valence gluons and correspond to hybrid mesons, which are energetically disfavored at low temperatures. Moreover, interactions at short distances in the octet channel are repulsive. 
Hence, we expect that the free energy is dominated by the singlet contribution for short enough distances: $ F_{\mathrm{avg}}(r,\beta,N_\tau) \approx F_S(r,T,N_\tau) +T\log 9 $. 
We account for the trivial color factor $ T\log 9 $ and show the modified continuum limit of the color-averaged free energy together with the static energy at zero temperature, $ V(r) $, in figure~\ref{fig: Facont}. 
Differences between $ F_{\mathrm{avg}}(r,T)-T\log 9 $ and $ V(r) $ due to thermal modification become significant for 
much smaller $ r $ due to the 
contribution from states in color-octet configuration. 
Results for the Polyakov loop correlator up to $ T \leq 350\,\mathrm{MeV} $ using stout-smeared staggered quarks that were presented in~\cite{Borsanyi:2015yka} are consistent with our results.

\subsection{Cyclic Wilson loop}\label{sec: Cyclic Wilson loop}

\enlargethispage{1\baselineskip}
 
\begin{figure}
\centering
  \begin{minipage}{0.48\textwidth}
\centering
 \includegraphics[height=4cm,clip]{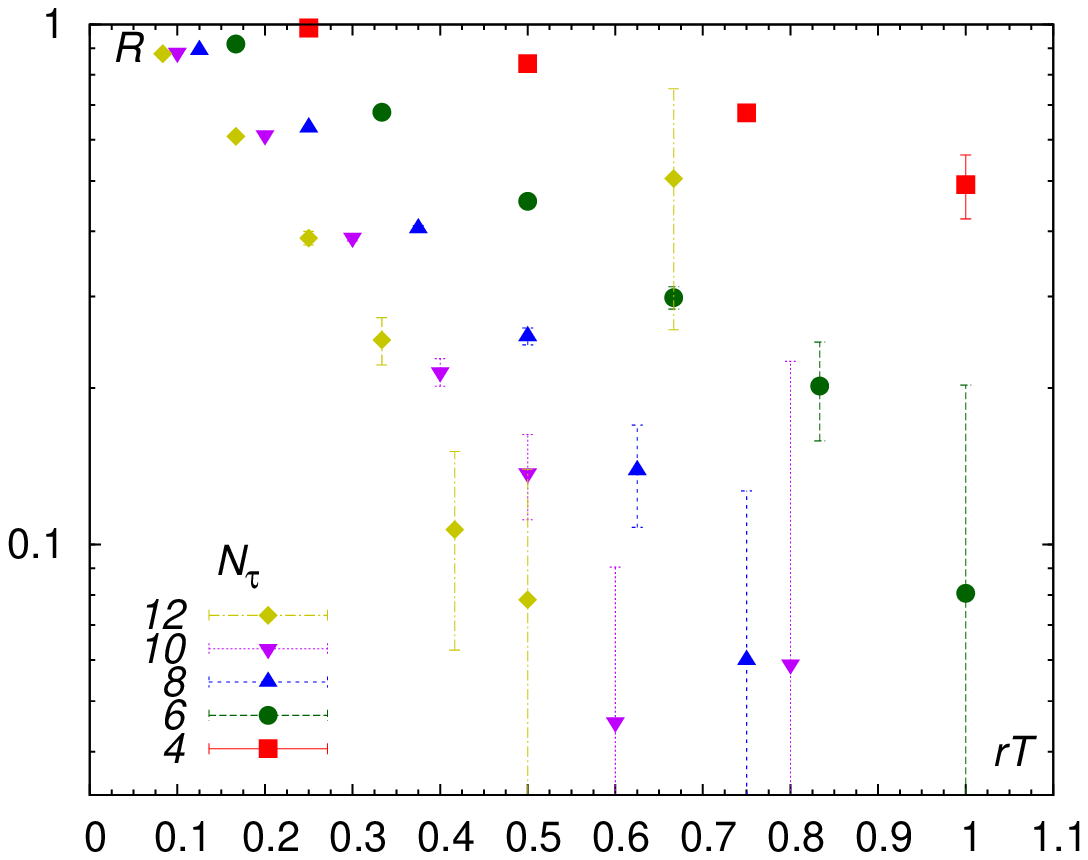}
  \vskip1ex
  \caption{\label{fig: ratio}
  The ratio $ R $ for $ T \approx 330\,\mathrm{MeV} $ exhibits the linear divergence.
  }
\end{minipage}\hfill
\begin{minipage}{0.48\textwidth}
\centering
\includegraphics[height=4cm,clip]{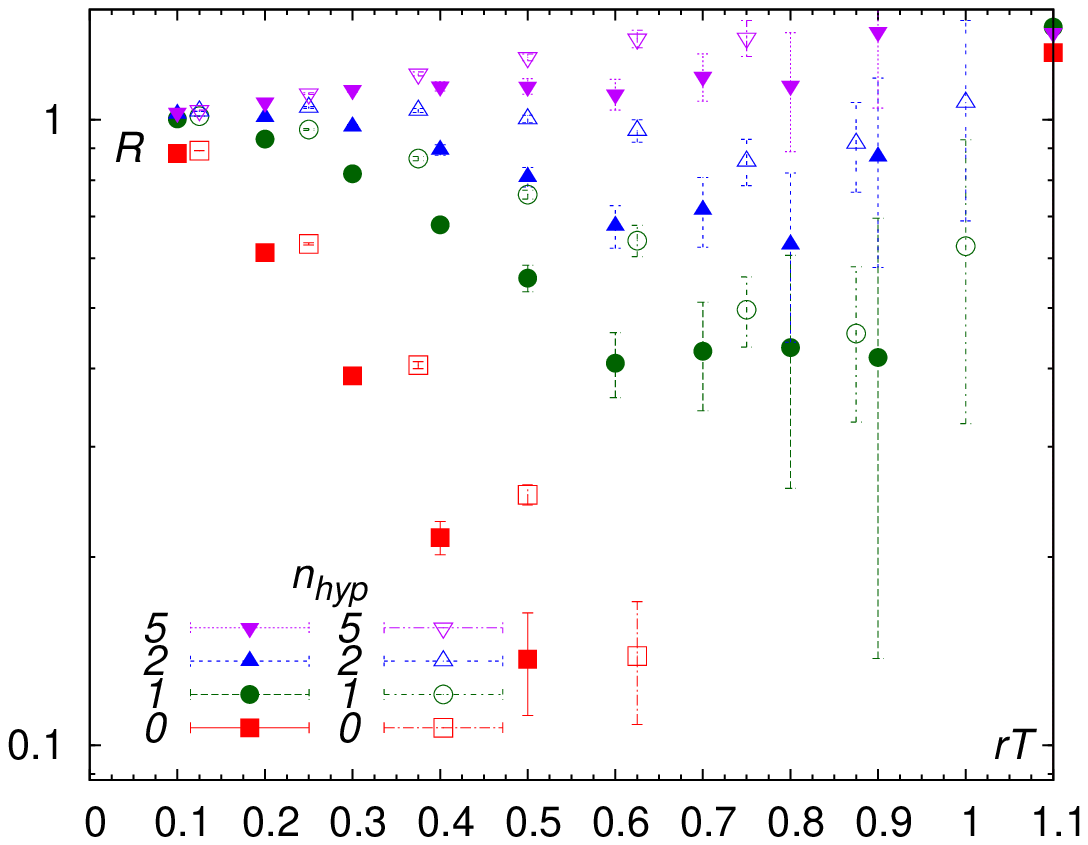}
  \caption{\label{fig: smearing}
  HYP smearing regulates the linear divergence. Different amounts $ n_{\mathrm{hyp}} $ are needed for $ N_\tau=8 $ (open symbols) and $ N_\tau=10 $ (filled symbols).
  }
\end{minipage}
\end{figure}

The cyclic Wilson loop and singlet or octet free energies defined in terms of the Wilson line correlator are not straightforwardly related due to both the linear divergence and finite differences that are analytic in $ rT $ and independent of $ N_\tau $. Motivated by \mbox{eq.}~(\ref{eq: multiplicative renormalization}), we take the ratio $ {R}(r,T,N_\tau) $, 
\begin{equation}
  {R}(T,r,N_\tau)= \frac{W_S(T,r,N_\tau)-C_P(T,r,N_\tau)}{C_S(T,r,N_\tau)-C_P(T,r,N_\tau)}.
\end{equation}
The divergence is apparent since $ R $ decreases exponentially. The rate of decrease grows with increasing $ N_\tau $ as in figure~\ref{fig: ratio}. 
We show in figure~\ref{fig: smearing} that the exponential decrease of $ R $ is greatly reduced by HYP smearing and $ R $ eventually even rises slightly above 1. 
Hence, a sufficiently smeared cyclic Wilson loops is a gauge invariant way to access static $ Q \bar Q $ free energies.

\section{Conclusions}
\vskip-1ex

\enlargethispage{1\baselineskip}

We have studied the Polyakov loop, its correlator, a color-singlet Wilson line correlator in Coulomb gauge and the cyclic Wilson loop in full QCD with 2+1 flavors of quarks almost at the physical point. 
We calculate free energies from these observables and study 
color screening.


The free energies of a static $ Q \bar Q $ pair in the medium are consistent with the $ T=0 $ static energy $ V(r) $ for short distances, indicating vacuum-like physics. 
For larger distances, the normalized color-averaged free energy $ F_{\mathrm{avg}}(r,T) -T\log 9 $ starts to deviate from $ V(r) $ at $ rT \approx 0.15 $ due to its octet contribution, whereas the color-singlet free energy $ F_S(r,T) $ stays close to $ V(r) $ up to $ rT \approx 0.45 $. We attribute this starting deviation to the onset of color screening, which is highlighted more explicitly in an effective coupling constant $ \alpha_{Q \bar Q}(r,T) $. 
The cyclic Wilson loop exhibits a linear divergence that can be regulated by smearing 
such that it becomes numerically similar to the Wilson line correlator.

\section*{Acknowledgments}
\vskip-1ex

This work was supported by U.S. Department of Energy under Contract No. DE-SC0012704. 
We acknowledge the support by the DFG Cluster of Excellence ``Origin and Structure of the Universe''. The calculations have been carried out on the computing facilities of the Computational Center for Particle and Astrophysics (C2PAP). The speaker thanks Matthias Berwein for many valuable discussions.

\end{document}